
\magnification=\magstep1
\abovedisplayskip=5pt plus 3pt minus 1pt
\belowdisplayskip=5pt plus 3pt minus 1pt
\belowdisplayshortskip=5pt plus 3pt minus 2pt
\hsize=5.75truein
\vsize=8.75truein
\baselineskip=12pt
\nopagenumbers
\centerline{RADIO AND OPTICAL EMISSION,}
\centerline{SPECTRAL SHAPES AND BREAKS IN GRB}
\bigskip
\centerline{J. I. Katz}
\centerline{Washington University, St. Louis, Mo. 63130}
\bigskip
\centerline{ABSTRACT}
\bigskip
Relativistic blast wave models of GRB predict the spectrum of the emitted
synchrotron radiation.  The electrons in the shocked region are heated to a
Wien distribution whose ``temperature'' is $1/3$ of the mean electron
energy.  This energy determines a characteristic (break) frequency of
synchrotron radiation.  At much lower frequencies a spectrum $F_\nu \propto
\nu^{1/3}$ is predicted independently of the details of the emitting region.
This is consistent with the observed soft X-ray emission of GRB.  It implies
low visible and radio intensities, unless there are collective emission
processes.
\bigskip
\centerline{INTRODUCTION}
\bigskip
Most of the controversies surrounding GRB involve, directly or indirectly,
the shapes of their continuum spectra.  Different theoretical models predict
different spectral characteristics.  For example, if the radiation emerges
from a stationary region optically thick to gamma-gamma pair production,
there will be a spectral break at an energy $O(m_e c^2)^1$.  If the source
emits as a black body the spectrum will resemble a Planck function, as may
be the case for SGR (but not for classical GRB).  Electrons with a power-law
distribution of energies produce optically thin synchrotron radiation with a
power-law spectrum and no characteristic energies or spectral breaks.

The low-frequency extension of the gamma-ray spectrum determines the
observability of GRB outside the gamma-ray band$^2$.  Observations at visible
frequencies$^3$ are widely believed to hold the key to identifying the
quiescent counterparts, astronomical sites, and physical mechanisms of GRB.
In addition, observations at soft X-ray frequencies$^4$ may provide a direct
measure of the intervening column density of (chiefly) oxygen, while
observations of radio dispersion similarly measure$^{5,6,7}$ the intervening
column density of free electrons.  These observations may settle the
question of Galactic {\it vs.}~cosmological distances for GRB, as well as
measure properties of the intergalactic medium if the distances are
cosmological.
\bigskip
\centerline{HYPOTHESIS}
\bigskip
A model of GRB has been developed which unambiguously predicts the shape of
the low-frequency part of their spectra.  This model involves debris$^8$
accelerated by a relativistic fireball interacting with a clumpy surrounding
interstellar medium$^{7,9,10}$; the observed gamma-rays are produced in
relativistically shock-heated interstellar matter and fireball debris.

In the more familiar case of relativistic particle acceleration at a
nonrelativistic shock (such as in supernova remnants) only a small fraction
of the particles are accelerated.  The acceleration process provides no
characteristic energy scale for the accelerated particles, so their spectrum
is a power law, broken only at the energy at which their gyroradii carry
them out of the region of acceleration.  This conventional model of shock
acceleration is inapplicable to relativistic shocks in GRB, contrary to
previous assertion$^7$.

In the present model of GRB all the charged particles in the shocked matter
are accelerated, and the internal energy per particle sets
an energy scale.  This internal energy is determined by the hydrodynamic
jump conditions at the shock, which in turn are set by the velocity of the
debris and the densities of the debris and interstellar medium.  Because
there is a characteristic energy scale (enforced by conservation of energy,
which is inapplicable to particles accelerated in an imposed flow field)
there is no reason to expect a power-law energy distribution of the
relativistic particles.  They rapidly interact with each other by means of
plasma waves (which mediate the collisionless shock), and come to an
equilibrium Wien distribution
$$N_e(E) \propto E^2 \exp(-E/k_B T), \eqno(1)$$
with the temperature parameter $k_B T$ being $1/3$ of the mean energy
$\cal E$ per particle.  Estimates show that the plasma wave interaction and
acceleration times, typically $O(\omega_g^{-1})$, where $\omega_g$ is the
gyrofrequency, are very much shorter than the other characteristic times in
the problem, the hydrodynamic rarefaction and the synchrotron radiation
times, so that these latter processes affect $\cal E$ but not the form (1).
\bigskip
\centerline{HIGH ENERGY SPECTRA}
\bigskip
The observed GRB spectra at high photon energies do not show the exponential
cutoff implied by the Wien particle spectrum (1).  There are two possible
explanations:
\item{1.} The radiating electrons interact with each other by means
of plasma waves which have a very high brightness temperature (far in excess
of the individual particle energies), and which therefore do not constitute
a genuine heat bath.  As a result, the form (1) is not thermodynamically
required, and a power-law spectrum (rather than an exponential cutoff) for
$E > {\cal E}$ is possible.  In order that this high energy tail not
dominate the energy content (which would be inconsistent with the definition
of $\cal E$) the electron energy distribution $N_e(E) \propto E^{-p}$ must
have an index $p > 2$ and the spectral index of its synchrotron radiation
(defined by $F_\nu \propto \nu^{-s}$) $s = (p - 1)/2 > 1/2$, consistent with
the observed$^2$ $s \approx 1$ at high energies.
\item{2.} At any time (and even more so in time-average) the observed
radiation is integrated over radiating volumes with a distribution, probably
very broad, of values of $\cal E$, magnetic field, and Doppler shift.  As a
result, the inferred distribution of energies of radiating particles only
shows an exponential cutoff at energies higher than the greatest $\cal E$
found anywhere in the radiating volume.  Observed breaks in the
spectrum$^{11,12}$ reflect a characteristic $\cal E$ in the radiating
region, and their evolution through a burst reflects the evolution of $\cal
E$ as the blast wave progresses through the interstellar medium.
\bigskip
\centerline{LOW ENERGY SPECTRUM}
\bigskip
For any electron energy distribution with power law exponent $p < 1/3$
synchrotron radiation at frequencies below the spectral peak is dominated by
the highest energy electrons because the power radiated at a given
frequency$^{13}$ is $\propto E^{1/3}$.  This condition is met by the Wien
distribution, for which $p \to -2$ for $E \ll {\cal E}$, and by most
plausible distributions below their characteristic energy $\cal E$.  The
integrated spectrum then has the index $s = - 1/3$:
$$F_\nu \propto \nu^{1/3}, \eqno(2)$$
characteristic of low-frequency synchrotron emission below the spectral
peak$^{13}$.  This result survives averaging over an emission region with a
range of electron energy distributions, $\cal E$, magnetic field, and
Doppler shift, as long as the frequency of observation is everywhere below
the characteristic synchrotron frequency (Doppler-shifted to the observer's
frame) for electrons of energy $\cal E$, and is therefore a robust
prediction of relativistic blast wave models.

The predicted spectrum (2) is consistent with data$^2$ on GRB at X-ray
energies below 10 KeV, where the observed photon count rate per unit energy
$N_\gamma \propto (h \nu)^{-0.7}$ is equivalent to $s = - 0.3$,
indistinguishable from $s = - 1/3$.  This
supports the applicability of relativistic blast wave models to GRB.  The
form (2) also resolves the X-ray paucity problem$^{14}$ which arises in
models of GRB emission close to neutron stars.

It is possible to extrapolate (2) to lower frequencies with confidence, once
the basic model is accepted.  An intense GRB with a flux $10^{-5}$
erg/cm$^2$sec in a soft gamma-ray bandwidth of 400 KeV has a gamma-ray flux
of 10 mJy (1~Jy~$\equiv 10^{-23}$ erg/cm$^2$ sec Hz) and a visible flux of 0.2
mJy.  The total visible power corresponds to a $\approx 18$th magnitude
star, difficult to detect as an optical transient.

For the same bright GRB extrapolation to 1 GHz leads to a predicted flux of
about 2 $\mu$Jy.  The effective flux of a brief transient measured by a
broad-band receiver may be further reduced by dispersion.  In addition,
self-absorption in an incoherent source leads to an independent upper
bound$^7$
$$F_\nu < 2 \pi \nu^2 m_p {r_s^2 \over D^2} \approx 0.5\ \mu{\rm Jy}
\left({\nu \over 10^9\ {\rm Hz}}\right)^2 \left({r_s \over 2 \times 10^{15}
\ {\rm cm}}\right)^2 \left({1\ {\rm Gpc} \over D}\right)^2, \eqno(3)$$
where $r_s$ is the radius of emission and $D$ is the distance to the GRB.
The self-absorption bound (3) will exceed the $\mu$Jy level for the
brightest GRB, which may be much closer than 1 Gpc, and after the observable
gamma-ray emission, when the blast wave expands to $r_s \gg 2 \times
10^{15}$ cm.

As the blast wave expands the number of radiating electrons increases
$\propto r^3$.  At a given frequency $\nu$ the radiated power density per
electron (in the co-moving frame in which the electron and field
distributions are assumed isotropic) is $\propto (\nu \gamma_e /B)^{1/3}$;
$\gamma_e$ and $B$ each vary$^7$ $\propto r^{-3/2}$.  Applying a Lorentz
factor $\gamma$ to transform to the observer's frame multiplies the spectral
density by $\gamma^{1 + s} = \gamma^{2/3} \approx \gamma_F^{1/3} \propto
r^{-1}$, where $\gamma_F$ is the fireball Lorentz factor.  Hence the
spectral brightness increases $\propto r^2 \propto t^{4/5}$ until a time
$t_c \propto \nu^{-5/12}$ characteristic of the frequency of observation$^7$
is reached, after which the brightness may fall exponentially; $t_c$ is
about $3 \times 10^4$ times longer at 1 GHz than at 300 KeV, and may be
days to weeks.  The peak brightness at 1 GHz may be $\sim 10^4$ times
brighter ($\sim 20$ mJy) than it was when the spectral peak was at 300 KeV.
These numerical results are necessarily rough.

The preceding arguments would have predicted that pulsars should be
unobservable at radio frequencies!  Fortunately, collective emission
processes occur which are not bound by single electron radiation rates or by
limits on brightness temperatures.  We may hope (with faint reason) that
collective processes are similarly effective in GRB.

I thank M. Davis, P. Horowitz, and B. E. Schaefer for discussions and NASA
NAGW 2918 for support.
\bigskip
\centerline{REFERENCES}
\bigskip
\item{1.} B. J. Carrigan and J. I. Katz {\it Ap. J.} {\bf 399}, 100 (1992).
\item{2.} B. E. Schaefer these proceedings.
\item{3.} B. E. Schaefer in {\it Gamma-Ray Bursts, Observations, Analyses
and Theories} eds.~C.~Ho, R.~I.~Epstein, E.~E.~Fenimore (Cambridge University
Press, Cambridge, 1992), p. 107.
\item{4.} B. E. Schaefer in {\it Compton Gamma-Ray Observatory} eds. M.
Friedlander, N. Gehrels, D. J. Macomb (AIP, New York, 1993), p. 803.
\item{5.} V. L. Ginzburg {\it Nature} {\bf 246}, 415 (1973).
\item{6.} D. M. Palmer {\it Ap. J. (Lett.)} in press (1993).
\item{7.} J. I. Katz {\it Ap. J.} in press (1994).
\item{8.} A. Shemi and T. Piran {\it Ap. J. (Lett.)} {\bf 365}, L55 (1990).
\item{9.} M. J. Rees and P. M\'esz\'aros {\it Mon. Not. Roy. Astr. Soc.}
{\bf 258}, 41p (1992).
\item{10.} P. M\'esz\'aros and M. J. Rees {\it Ap. J.} {\bf 405}, 278
(1993).
\item{11.} B. E. Schaefer, {\it et al.} {\it Ap. J. (Lett.)} {\bf 393}, L51
(1992).
\item{12.} R. D. Preece these proceedings.
\item{13.} J. D. Jackson {\it Classical Electrodynamics} (Wiley, New York,
1975).
\item{14.} J. N. Imamura and R. I. Epstein {\it Ap. J.} {\bf 313}, 711
(1987).
\vfil
\eject
\bye
\end